# Anisotropic Bianchi Type-$V$ Bulk Viscous Universe with Variable $G$ and $\Lambda$

Nawsad Ali

*Department of Mathematics,*

*West Goalpara College, Balarbhita-783129, Goalpara, Assam, India*

Email: alinawsad@rocketmail.com

**Abstract**: The present study deals with spatially homogeneous and totally anisotropic Bianchi type-$V$ cosmological model in presence of bulk viscous fluid source with time dependent gravitational constant $G$ and cosmological term $\Lambda$. The coefficient of bulk viscosity $\zeta$ is considered to be a quadratic function of Hubble parameter $H$ (i.e. $\zeta = \zeta_0 + \zeta_1 H + \zeta_2 H^2$, where $\zeta_0$, $\zeta_1$ and $\zeta_2$ are constants). Exact solution of Einstein's field equations are obtained by using a particular form of Hubble's parameter $H$ i.e. $H(a) = m(a^{-n} + 1)$, where $m > 0$, $n > 1$, $a$ being a scale factor and $H = \dfrac{\dot{a}}{a}$, which provides an early deceleration and late time acceleration of the universe. From state finder diagnosis $\{r, s\}$ the behavior of the universe in different stages during the evolution is exhibits.

## 1. Introduction

From the recent Supernovae observations [1-3], the Maxima [4], Boomerang [5] data on cosmic microwave background radiation (CMBR) and from WMAP data [6, 7] it is clear that the present universe is expanding with ever accelerating. This is cause due to an effective negative pressure and for this there are some possible candidates dubbed as dark energy [8-10]. Among many possible alternatives, the simplest and most theoretically appealing for dark energy is the energy density stored on the vacuum state of all existing fields in the universe, i.e., $\rho_v = \dfrac{\Lambda}{8\pi G}$, where $\Lambda$ is the cosmological constant. However, a constant $\Lambda$ cannot explain the huge difference between the cosmological constant inferred from observation and the vacuum energy density resulting from quantum field theory. To accommodate and interpret this problem, variable $\Lambda$ especially time dependent was pointed out so that $\Lambda$ was large in the early universe and then decayed with evolution [11]. In recent past, cosmological models with different decay laws for the variation of cosmological term were investigated by several researchers [12-20].

The concept of a variable gravitational constant $G$ was first proposed by Dirac [21]. Further, Canuto and Narlikar [22] have analyzed that $G$ varying cosmology consistent with whatsoever cosmological observations available at present epoch. In addition, Lau [23] has suggested the modification linking the variation of $G$ with that of $\Lambda$ in the framework of general relativity. Therefore, this modification allows us the form of Einstein's field equations formally unchanged since a variation of $G$ together with the variation of $\Lambda$. In recent times, linking the variation of $G$ with that of $\Lambda$ in the framework of relativistic cosmology have been discussed by several authors [24-30].

The investigations of the most cosmological models usually assume that the matter in the universe may be described by a perfect fluid. A realistic treatment of the problem requires the consideration of material distribution other than the perfect fluid. It is believed that during neutrino decoupling stage, apart from streaming neutrinos moving with fundamental velocity, the matter behaved like a viscous fluid in the early stages of evolution [31]. Also, decoupling of radiation and matter during the recombination era is also expected to give rise to viscous effects. In addition, a combination of cosmic fluid together with bulk dissipative pressure can generate accelerate expansion of the universe [32]. The bulk viscous effect leading to an accelerated phase of the universe has been explained by Fabris et al. [33]. The effect of viscosity on the evolution of the cosmological models has been suggested by Misner [34].

In general the bulk viscosity is taken as a time dependent in literature. The coefficient of viscosity decreases as universe expands. Padmanabhan and Chitre [35] have shown the effect of bulk viscosity on the evolution of the universe at large. Bulk viscosity assumed to be power functions of matter density has been claimed by Singh [36] in LRS Bianchi type-$V$ space-time in the presence of imperfect fluid accompanied with heat conduction. Cosmological models in presence of bulk viscous fluid, the viscous coefficient can be considered as a linear and quadratic function of Hubble parameter $H$ has been presented by Meng and Ma [37]. Later on Baghel and Singh [38] have analyzed Bianchi type-$V$ bulk viscous universe with time function $G$ and $\Lambda$, where the coefficient of bulk viscosity chosen as a linear function of Hubble parameter $H$ (i.e., $\zeta = \zeta_0 + \zeta_1 H$). In similar way, Das and Ali [39] have discussed axially symmetric Bianchi type-$I$ bulk viscous cosmological models with time dependent $G$ and $\Lambda$, where they have considered coefficient of bulk viscosity as a quadratic function of Hubble parameter $H$ (i. e., $\zeta = \zeta_0 + \zeta_1 H + \zeta_2 H^2$). Recently, the effect of bulk viscosity on cosmological evolution in the framework of general theory of relativity have been studied by many researchers [40-42].

The simplest model of the observed universe is well represented by Friedmann-Robertson-Walker (FRW) models, which are both spatially homogeneous and isotropic. However, these models in some sense are good global approximation of the present day universe. But on smaller scales, the universe is neither homogeneous nor isotropic. In case of theoretical arguments it is believed that the early universe was characterized by a highly irregular mechanism which isotropized later [43, 44]. Anisotropic and spatially-homogeneous cosmological models provide a richer structure both geometrically and physically [45]. Bianchi type-$V$ models being anisotropic generalization of open FRW models are interesting to study. These models are favoured by the available evidences for low density of universe. Coley [46] discussed Bianchi type-$V$ viscous fluid cosmological models for barotropic fluid distribution. A self consistent system of gravitational field with a binary mixture of perfect fluid and dark energy given by a cosmological constant has been considered in anisotropic Bianchi type-$V$ universe by Singh and Chaubey [47]. One year later, they (Singh and Chaubey [48]) have analyzed the evolution of a homogeneous anisotropic Bianchi type-$V$ universe field with viscous fluid. Bianchi type-$V$ cosmological models have been studied by several authors in different context [49-53].

In this paper we extent the earlier work done by Baghel and Singh [38] in connection to anisotropic Bianchi type-$V$ universe in the presence of bulk viscous fluid with time function $G$ and $\Lambda$. Here the coefficient of bulk viscosity is supposed to be a quadratic function of Hubble parameter $H$ (i.e. $\zeta = \zeta_0 + \zeta_1 H + \zeta_2 H^2$, where $\zeta_0$, $\zeta_1$ and $\zeta_2$ are constants). Following Ellis and Madsen [54], to get the deterministic model of the universe, we also assumed a specific form of Hubble parameter $H$ i.e., $H(a) = m(a^{-n} + 1)$, where $m > 0$, $n > 1$, $a$, being a scale factor and $H = \frac{\dot{a}}{a}$, which gives an early deceleration and current time acceleration of the universe.

This paper is organized as follows. In Sect. 2, we describe our model and field equations. A particular form of generalized Hubble's parameter, which gives an early deceleration and current time acceleration and the coefficient of bulk viscosity as a quadratic function of Hubble parameter are presented in Sect. 3. In Sect. 4 we discussed the solutions of the field equations. In Sect. 5 we discussed some physical and geometrical aspects. Sect. 6 deals with statefinder diagnostic. Finally, we summarized our conclusions in the last Sect. 7.

## 2. The Metric and Field Equations

The line element for Bianchi type-$V$ space time in an orthogonal form is given by

$$ds^2 = -dt^2 + A^2(t)dx^2 + e^{2\alpha x}\{B^2(t)dy^2 + C^2(t)dz^2\}, \tag{1}$$

where $\alpha$ is a constant. We consider the cosmic matter consisting of bulk viscous fluid represented by the energy momentum tensor

$$T_{ij} = (\rho + \bar{p})v_i v_j + \bar{p} g_{ij}, \tag{2}$$

with

$$\bar{p} = p - \zeta v^i_{;i}, \tag{3}$$

satisfying a linear equation of state

$$p = \omega \rho, \tag{4}$$

where $p$ is the equilibrium pressure, $\rho$ is the energy density of matter, $\zeta$ is the coefficient of bulk viscosity and $v^i$ is the flow vector of the fluid satisfying $v_i v^i = -1$. The semicolon in equation (3) stands for covariant differentiation. On thermodynamical grounds bulk viscous coefficient $\zeta$ is positive, assuring that the viscosity pushes the dissipative pressure $\bar{p}$ towards negative values. The Einstein's field equations with time varying $G(t)$ and $\Lambda(t)$ are given by

$$R_i^j - \frac{1}{2} g_i^j R = -8\pi G(t) T_i^j + \Lambda(t) g_i^j. \qquad (5)$$

For the Bianchi type-$V$ space time with bulk viscous fluid distribution the Einstein's field equations (5) yield's the following equations

$$8\pi G \bar{p} - \Lambda = \frac{\alpha^2}{A^2} - \frac{\ddot{B}}{B} - \frac{\ddot{C}}{C} - \frac{\dot{B}\dot{C}}{BC}, \qquad (6)$$

$$8\pi G \bar{p} - \Lambda = \frac{\alpha^2}{A^2} - \frac{\ddot{C}}{C} - \frac{\ddot{A}}{A} - \frac{\dot{C}\dot{A}}{CA}, \qquad (7)$$

$$8\pi G \bar{p} - \Lambda = \frac{\alpha^2}{A^2} - \frac{\ddot{A}}{A} - \frac{\ddot{B}}{B} - \frac{\dot{A}\dot{B}}{AB}, \qquad (8)$$

$$8\pi G \rho + \Lambda = -\frac{3\alpha^2}{A^2} + \frac{\dot{A}\dot{B}}{AB} + \frac{\dot{B}\dot{C}}{BC} + \frac{\dot{C}\dot{A}}{CA}, \qquad (9)$$

$$0 = \frac{2\dot{A}}{A} - \frac{\dot{B}}{B} - \frac{\dot{C}}{C}, \qquad (10)$$

where an overhead dot (.) denotes ordinary differentiation with respect to cosmic time $t$. Vanishing divergence of Einstein tensor $R_i^j - \frac{1}{2} g_i^j R$ gives rise to

$$8\pi G \left\{ \dot{\rho} + (\rho + \bar{p}) \left( \frac{\dot{A}}{A} + \frac{\dot{B}}{B} + \frac{\dot{C}}{C} \right) + \frac{\dot{G}}{G} \rho \right\} + \dot{\Lambda} = 0. \qquad (11)$$

The usual energy conservation equation $T_{i;j}^j = 0$ yields

$$\dot{\rho} + (\rho + \bar{p}) \left( \frac{\dot{A}}{A} + \frac{\dot{B}}{B} + \frac{\dot{C}}{C} \right) = 0. \qquad (12)$$

Equation (11) together with equation (12) put $G$ and $\Lambda$ in some sort of couple field given by

$$8\pi \dot{G}\rho + \dot{\Lambda} = 0. \tag{13}$$

From (13) one concludes that, when $\Lambda$ is constant or $\Lambda=0$, $G$ turns out to be constant for non zero energy density. We define the average scale factor $a$ for the metric (1) as

$$a^3 = \sqrt{-g} = ABC. \tag{14}$$

From equations (6)-(9) and (10), we obtain

$$\frac{\dot{A}}{A} = \frac{\dot{a}}{a}, \tag{15}$$

$$\frac{\dot{B}}{B} = \frac{\dot{a}}{a} - \frac{k_1}{a^3}, \tag{16}$$

and

$$\frac{\dot{C}}{C} = \frac{\dot{a}}{a} + \frac{k_1}{a^3}, \tag{17}$$

with $k_1$ being a constant of integration. On integrating equations (15)-(17), we obtain

$$A = x_1 a, \tag{16}$$

$$B = x_2 a \exp\left\{-k_1 \int \frac{dt}{a^3}\right\} \tag{17}$$

and

$$C = x_3 a \exp\left\{k_1 \int \frac{dt}{a^3}\right\}, \tag{18}$$

where $x_1$, $x_2$ and $x_3$ are constants of integrations satisfying $x_1 x_2 x_3 = 1$.

In analogy with FRW universe, we define a generalized Hubble parameter $H$ and generalized deceleration parameter $q$ as

$$H = \frac{\dot{a}}{a} = \frac{1}{3}(H_1 + H_2 + H_3), \qquad (19)$$

$$q = -1 - \frac{\dot{H}}{H^2}, \qquad (20)$$

where $H_1 = \frac{\dot{A}}{A}$, $H_2 = \frac{\dot{B}}{B}$ and $H_3 = \frac{\dot{C}}{C}$ are directional Hubble's factors along $x$, $y$ and $z$ directions respectively.

We introduced the volume expansion $\theta$ and shear scalar $\sigma$ as usual

$$\theta = v^i_{;i} \; ; \quad \sigma^2 = \frac{1}{2}\sigma_{ij}\sigma^{ij}, \qquad (21)$$

$\sigma^{ij}$ being shear tensor and semicolon stand for covariant differentiation. For Bianchi type-$V$ model, expansion scalar $\theta$ and shear scalar $\sigma$ are given by

$$\theta = 3\frac{\dot{a}}{a}, \qquad (22)$$

$$\sigma = \frac{k_1}{a^3}. \qquad (23)$$

Equations (6)-(10) and (12) can be written in terms of $H$, $\sigma$ and $q$ as

$$8\pi G \bar{p} - \Lambda = (2q-1)H^2 - \sigma^2 + \frac{\alpha^2}{a^2}, \qquad (24)$$

$$8\pi G \rho + \Lambda = 3H^2 - \sigma^2 - \frac{3\alpha^2}{a^2}, \qquad (25)$$

and

$$\dot{\rho} + 3(\rho + \bar{p})H = 0. \qquad (26)$$

It is note that energy density of the universe is a positive quantity. It is believed that at the early stages of the evolution when the average scale factor $a$ was close to zero, the energy density of the universe was infinitely large. On the other hand, with the expansion of the universe i.e. with increase of $a$, the energy density decreases and an infinitely large $a$ correspond to $\rho$ close to zero. In that case from (25), we obtain $\dfrac{\rho_v}{\rho_c} \to 1$, where $\rho_v = \dfrac{\Lambda}{8\pi G}$ and $\rho_c = \dfrac{3H^2}{8\pi G}$. For $\Lambda \geq 0$, $\rho \leq \rho_c$. Also, from (25) we observe that

$$\frac{\sigma^2}{\theta^2} = \frac{1}{3} - \frac{8\pi G \rho}{\theta^2} - \frac{3\alpha^2}{a^2\theta^2} - \frac{\Lambda}{\theta^2}. \tag{27}$$

Therefore, $0 \leq \dfrac{\sigma^2}{\theta^2} \leq \dfrac{1}{3}$ and $0 \leq \dfrac{8\pi G \rho}{\theta^2} \leq \dfrac{1}{3}$ for $\Lambda \geq 0$. Thus the presence of a positive $\Lambda$ puts restriction on the upper limit of anisotropy where as a negative $\Lambda$ contributes to the anisotropy. From equation (24) and (25), we obtain

$$\frac{d\theta}{dt} = \Lambda + 12\pi\theta\zeta G - 4\pi G(\rho + 3p) - 2\sigma^2 - \frac{1}{3}\theta^2, \tag{28}$$

which is the Raychaudhuri equation for the given distribution. We observe that for $\Lambda \leq 0$ and $\zeta = 0$, the universe will always be in decelerating phase provided the strong energy condition (Hawking and Ellis [55]) holds, whereas in the presence of viscosity, positive $\Lambda$ will be slow down the rate of decrease of volume expansion. Also, $\dot{\sigma} = -\sigma\theta$ implies that $\sigma$ decrease in an evolving universe and for an infinitely large value of $a$, $\sigma$ becomes negligible.

### 3. Basic Assumptions of the Model

There are six linearly independent equations (4) and (6)-(10) with eight unknown parameters $A$, $B$. $C$, $p$, $\rho$, $G$, $\Lambda$ and $\zeta$. The system is thus initially undetermined together with the energy conservation relation (11), and we need two additional constraints to close the system. For any physically relevant model, the Hubble parameter $H$ and deceleration parameter $q$ are most important observational quantities in cosmology. The present day value $H_0$ of Hubble parameter sets the present time scale of the expansion while $q_0$, the present day value of deceleration parameter tells that expansion of the present universe is accelerating rather than going to decelerate as expected before the supernovae of type 1a observations. However, a number of recent observational data suggest that this accelerating phase of the universe is a recent phenomenon. The value of deceleration parameter separate the decelerating ($q > 0$) from accelerating ($q < 0$) periods in the evolution of the universe. Determination of the deceleration parameter from the count-magnitude relation for galaxies is a difficult task due to evolutionary effects. In order to an expanding model of the universe consistent with recent observations, one needs a Hubble parameter $H$ such that the model starts with a decelerating expansion followed by an accelerated expansion at late times. We also use well known relation investigated by, Ellis and Madsen [54], firstly, we consider a functional form of average Hubble parameter $H$, which despite being quite simple, will describe both decelerating and accelerating phase of the universe. The average Hubble parameter $H$ is related to the average scale factor $a$ of an anisotropic Bianchi type-$V$ space-time as

$$H(a) = m(a^{-n} + 1), \qquad (29)$$

where $m(>0)$ and $n(>1)$ are constants. For the choice, the deceleration parameter $q$ turns out to be

$$q = \frac{n}{a^n + 1} - 1. \qquad (30)$$

A similar form for $q$ has also been proposed by Banerjee and Das [56] and Singha and Debnath [57] in case of FRW space time and Singh [58, 59] and Bali and Singh [60] for Bianchi type space time. From equation (30), it is clear that when $a = 0$, $q = n - 1 > 0$; $q = 0$ for $a^n = n - 1$ and for $a^n > n - 1$, $q < 0$. We consider the value $a = 0$ for $t = 0$. Hence, the universe begins to expand with an acceleration and this expansion changes from decelerating phase to an accelerating one. This scenario is consistent with recent supernova observation (SN 1a).

Secondly, to determine the coefficient of bulk viscosity $\zeta$ is considered to be a quadratic function of Hubble parameter $H$, i.e.

$$\zeta = \zeta_0 + \zeta_1 H + \zeta_2 H^2, \qquad (31)$$

where $\zeta_0 (\geq 0)$, $\zeta_1$ and $\zeta_2$ are constants. The motive behind assuming this condition has explained by (Meng and Ma [37]. For this choice, equation (26) becomes to

$$\dot{\rho} + 3(1 + \omega)\rho H = 9\zeta H^2. \qquad (32)$$

### 4. Solution of the Field Equation

For this model average scale factor $a$ is given by

$$a^n = e^{mn(t+t_1)} - 1. \qquad (33)$$

where $t_1$ is a constant of integration. Setting $a = 0$ for $t = 0$, we get $t_1 = 0$. Therefore,

$$a^n = e^{mnt} - 1. \qquad (34)$$

Expansion scalar $\theta$, shear $\sigma$ and deceleration parameter $q$ for the model are

$$\theta = \frac{3m}{(1-e^{-mnt})}, \tag{35}$$

$$\sigma = \frac{k_1}{(1-e^{-mnt})^{\frac{3}{n}}}, \tag{36}$$

$$q = \frac{n}{e^{mnt}} - 1. \tag{37}$$

The expressions for the coefficient of bulk viscosity $\zeta$ and matter density $\rho$ are obtain as

$$\zeta = \zeta_0 + \frac{\zeta_1 m}{(1-e^{-mnt})} + \frac{\zeta_2 m^2}{(1-e^{-mnt})^2}, \tag{38}$$

$$\rho = \frac{3m(\zeta_0 + \zeta_1 m + \zeta_2 m^2)}{(1+\omega)} + \frac{9m(\zeta_0 + \zeta_1 m + 3\zeta_2 m^2)}{(3+3\omega-n)(e^{mnt}-1)} + \frac{9m^2(\zeta_1 + \zeta_2 m)}{(3+3\omega-2n)(e^{mnt}-1)^2}$$

$$+ \frac{9\zeta_2 m^3}{(3+3\omega-3n)(e^{mnt}-1)^3} + \frac{k_2}{\left(e^{mnt}-1\right)^{\frac{3(1+\omega)}{n}}}, \tag{39}$$

where $k_2$ is a constant of integration. The expressions of Newtonian gravitational constant $G$ and cosmological term-$\Lambda$ obtain as

$$G = \frac{1}{4\pi}\left[\frac{m^2 n e^{mnt}}{\left(e^{mnt}-1\right)^2} - \frac{k_1^2}{\left(e^{mnt}-1\right)^{\frac{6}{n}}} - \frac{\alpha^2}{\left(e^{mnt}-1\right)^{\frac{2}{n}}}\right] \times$$

$$\left[3m(\zeta_0 + \zeta_1 m + \zeta_2 m^2) + \left\{\frac{9m(1+\omega)(\zeta_0 + \zeta_1 m + 3\zeta_2 m^2)}{(3+3\omega-n)(e^{mnt}-1)} - 3\zeta_0 m^2 e^{mnt}\right\}\frac{1}{\left(e^{mnt}-1\right)}\right.$$

$$+ \left\{\frac{9m^2(1+\omega)(\zeta_1 + 3\zeta_2 m)}{(3+3\omega-2n)(e^{mnt}-1)} - 3\zeta_1 m^2 e^{2mnt}\right\}\frac{1}{\left(e^{mnt}-1\right)^2}$$

$$+ \left\{\frac{9m^2(1+\omega)(\zeta_1 + 3\zeta_2 m)}{(3+3\omega-2n)(e^{mnt}-1)} - 3\zeta_1 m^2 e^{2mnt}\right\}\frac{1}{\left(e^{mnt}-1\right)^2}$$

$$+ \left\{ \frac{9(1+\omega)\zeta_2 m^3}{(3+3\omega-3n)(e^{mnt}-1)} - 3\zeta_2 m^3 e^{3mnt} \right\} \frac{1}{(e^{mnt}-1)^3} + \frac{k_2(1+\omega)}{(e^{mnt}-1)^{\frac{3(1+\omega)}{n}}} \right]^{-1}, \qquad (40)$$

and

$$\Lambda = \left[ \frac{2k_1^2}{(e^{mnt}-1)^{\frac{6}{n}}} + \frac{2\alpha^2}{(e^{mnt}-1)^{\frac{2}{n}}} - \frac{2m^2 n e^{mnt}}{(e^{mnt}-1)^2} \right] \times \left[ \frac{3m(\zeta_0 + \zeta_1 m + \zeta_2 m^2)}{(1+\omega)} \right.$$

$$+ \frac{9m(\zeta_0 + \zeta_1 m + 3\zeta_2 m^2)}{(3+3\omega-n)(e^{mnt}-1)} + \frac{9m^2(\zeta_1 + \zeta_2 m)}{(3+3\omega-2n)(e^{mnt}-1)^2} + \frac{9\zeta_2 m^3}{(3+3\omega-3n)(e^{mnt}-1)^3} + \frac{k_2}{(e^{mnt}-1)^{\frac{3(1+\omega)}{n}}} \right]$$

$$\times \left[ 3m(\zeta_0 + \zeta_1 m + \zeta_2 m^2) + \left\{ \frac{9m(1+\omega)(\zeta_0 + \zeta_1 m + 3\zeta_2 m^2)}{(3+3\omega-n)(e^{mnt}-1)} - 3\zeta_0 m^2 e^{mnt} \right\} \frac{1}{(e^{mnt}-1)} \right.$$

$$+ \left\{ \frac{9m^2(1+\omega)(\zeta_1 + 3\zeta_2 m)}{(3+3\omega-2n)(e^{mnt}-1)} - 3\zeta_1 m^2 e^{2mnt} \right\} \frac{1}{(e^{mnt}-1)^2}$$

$$+ \left\{ \frac{9m^2(1+\omega)(\zeta_1 + 3\zeta_2 m)}{(3+3\omega-2n)(e^{mnt}-1)} - 3\zeta_1 m^2 e^{2mnt} \right\} \frac{1}{(e^{mnt}-1)^2}$$

$$+ \left\{ \frac{9(1+\omega)\zeta_2 m^3}{(3+3\omega-3n)(e^{mnt}-1)} - 3\zeta_2 m^3 e^{3mnt} \right\} \frac{1}{(e^{mnt}-1)^3} + \frac{k_2(1+\omega)}{(e^{mnt}-1)^{\frac{3(1+\omega)}{n}}} \right]^{-1}$$

$$+ \frac{3m^2 e^{2mnt}}{(e^{mnt}-1)^2} - \frac{k_1^2}{(e^{mnt}-1)^{\frac{6}{n}}} - \frac{3\alpha^2}{(e^{mnt}-1)^{\frac{2}{n}}}. \qquad (41)$$

## 5. Physical and Geometrical Properties

We observe that spatial volume of the model is zero at $t=0$. Therefore, the present model starts evolving at $t=0$ and expands with cosmic time $t$. The model has singularity at $t=0$. At $t=0$, the values of $\zeta$, $\rho$, $\Lambda$, $G$, $\theta$, $\sigma$ all are infinite and deceleration parameter $q = n-1 \, (>0)$. Thus, the expansion in the model decelerates. At $t \to \infty$, $\theta$, $\sigma$ become zero, $\zeta = \zeta_0$ and $\rho = \frac{3m(\zeta_0 + \zeta_1 m + \zeta_2 m^2)}{(1+\omega)}$. We observe from equation (39) the matter density is a decreasing function of cosmic time as shown in the Fig. 1. From equation (40) we have plotted $G$ with respect to cosmic time $t$, which is negative in nature (see Fig. 2). The possibility of a negative gravitational constant $G$ has been discussed by Starobinskii [61], who concluded that the effective gravitational constant may have changed sign in the early universe. From equation (41), we have seen that the cosmological term $\Lambda$ being very large at initial times relaxes to a genuine cosmological constant at the late times (see Fig. 3). Also, for large values of $t$ i.e., $t \to \infty$, we obtain $q = -1$, which characterizes the de Sitter solution (Weinberg [62]). Therefore the model starts with a decelerating expansion and expansion in the model changes from decelerating phase to an accelerating one. Time $t_q$ when the expansion changes from the decelerating to accelerating phase can be obtained by equating $q = 0$ in (30) and it comes out to be

$$t_q = \frac{\ln n}{mn}. \qquad (42)$$

For the present model,

$$\frac{\sigma}{\theta} = \frac{k_1(1-e^{-mnt})}{3m(e^{mnt}-1)^{\frac{3}{n}}}. \qquad (43)$$

Since $\lim_{t \to \infty} \frac{\sigma}{\theta} = 0$, hence the model isotropizes for large values of $t$. For this solution, the age of the universe is given by

$$mnt_0 = In\left(\frac{H_0}{H_0 - m}\right), \qquad (44)$$

where $H_0$ denotes the present day value of $H$.

**6. Statefinder Diagnostic**

Sahni et al. [63] and Alam et al. [64] introduced a pair of parameter $\{r, s\}$, called statefinder parameters. This statefinder parameters can effectively differentiate between different form of dark energy and provide simple diagnosis regarding whether a particular model fits into the basic observation data. The above statefinder diagnostic pair has the following form

$$r = 1 + 3\frac{\dot{H}}{H^2} + \frac{\ddot{H}}{H^3} \text{ and } s = \frac{r-1}{3\left(q - \frac{1}{2}\right)}, \qquad (45)$$

For our bulk viscous model, parameter $\{r, s\}$, can be explicitly written in terms of $a$ as

$$r = 1 + \frac{n^2 - 3n}{1 + a^n} + \frac{n^2}{(1 + a^n)^2}, \tag{46}$$

$$s = \frac{2(n^2 - 3n)(1 + a^n) + 2n^2}{(1 + a^n)\{6n - 9(1 + a^n)\}}, \tag{47}$$

From (46) and (47) the complicated form between $r$ and $s$ obtained as

$$\left(\frac{6sn}{9s + 2r - 2}\right)^2 = \frac{6sn(n^2 - 3n)}{(r-1)(9s + 2r - 2)} + \frac{n^2}{r - 1}. \tag{48}$$

From equation (48) we plot Fig. 4 which reveals the variation of $s$ with the variation of $r$ for $n = 2$. The positive side of the curve $s$ starts from radiation era and goes asymptotically to the dust era ($s \to \infty$) whereas the negative side of the curve $s$ represents the evolution starts from dust era to $\Lambda CDM$ model ($r = 1$, $s = 0$) and then the region corresponding to $r < 1$, $s > 0$ the universe entered the phase of quintessence era. Therefore, the complete curve refers the universe start from radiation era to the $\Lambda CDM$ and then quintessence model..

## 7. Conclusion

In this paper, we have presented, Einstein's field equations with time varying gravitational constant $G$ and cosmological term $\Lambda$ in the context of Bianchi type-$V$ space time and bulk viscous fluid source. The coefficient of bulk viscosity is assumed to be a quadratic function of Hubble parameter $H$ (i.e. $\zeta = \zeta_0 + \zeta_1 H + \zeta_2 H^2$, where $\zeta_0$, $\zeta_1$ and $\zeta_2$ are constants). The effect of bulk viscosity is to introduce a change in the perfect fluid model. The spatial volume increases as time increases. For $\zeta_0 = \zeta_1 = \zeta_2 = 0$ our presents model becomes to a perfect fluid model. It is observed that the cosmological term is decreasing function of cosmic time (see Fig. 3) and approaches to zero at late time which is supported by recent result from the observations of type 1a supernova explosion.. The model asymptotically tends to a de-Sitter universe for large values of $t$, the model becomes isotropic which is in accordance with the high isotropy of the cosmic background radiation. The role of bulk viscosity is also exhibited in this model of the universe The proposed specific form of Hubble's parameter leads to a cosmological scenario in accordance with well based feature of modern cosmology in which initial stage is dominated by decelerating phase followed by an epoch at late time in which universe expands with acceleration. Our model presents a unified description of the evolution of the universe which starts with a deceleration expansion and expands with acceleration at late time as supported by observational data (Perlmutter et al. [2]; Knop et al. [65]; Tegmark et al. [66]; Spergel et al. [67].

Finally, based on the behavior of the evolutionary of the statefinder trajectory which is presented in Fig. 4 we can conclude that the model meets the $\Lambda CDM$ limit and then enter the phase of quintessence era. The above conclusion shows that our model is in good agreement with the recent scenario to describe the dynamics of universe at the present epoch.

**References**

[1]     Perlmutter, S., et al.: Astrophys. J. **483**, 565 (1997)

[2]     Perlmutter, S. et al.: Nature **391**, 51 (1998)

[3]     Riess, A. G., et al.: Astron. J. **116**, 1009 (1998)

[4]     Lange, A. E., et al.: Phys. Rev. D **63**, 042001 (2001)

[5]     Balbi, A., et al.: Astrophys. J **545**, L1 (2000)

[6]     Spergel, D. N. et al.: Astrophys. J. Suppl. **148**, 175 (2003)

[7]     Spergel, D. N. et al.: Astrophys. J. Suppl. **170**, 377 (2007)

[8]     Sahni, V., Starobinsky, A. A.: Int. J. Mod. Phys. A, **9,** 373 (2000)

[9]     Peebles, P. J. E., Ratra, B.:Rev. Mod. Phys. **75**, 559 (2003)

[10]    Padmanabhan, T.: Phys. Rep. **380**, 235 (2003)

[11]    Dolgov, A. D.: In The Very Early Universe (eds. Gibbons, G. W., Hawking, S. W., Siklos, S. T. C.), Cambridge University Press, Cambridge (1983)

[12]    Ratra, B., Peebles, P. J. E: Phys. Rev. D, **37**, 3406 (1988)

[13]    Chen, W., Wu, Y. S.: Phys. Rev. D **41**, 695 (1990)

[14]    Pavon, D.: Phys. Rev. D **43**, 375 (1991)

[15]    Carvalho, J. C., Lima, J. A. S., Waga, I.: Phys. Rev. D **46**, 2404 (1992)

[16]    Lima, J. A. S., Maia, J. M. F.: Phys. Rev. D **49**, 5579 (1994)

[17]    Lima, J. A. S., Trodden, M.: Phys. Rev. D **53**, 4280 (1996)

[18]    Arbab, A. I., Abdel-Rahman, A.-M. M.: Phys. Rev. D **50**, 7725 (1994)

[19]    Vishwakarma, R. G.: Gen. Relativ. Grav. **33**, 1973 (2001)

[20]   Ali, N.: J. Astrophys. Astr. **34** 259 (2013)

[21]   Dirac, P. A. M.: Proc. R. Soc. **165**, 199 (1938)

[22]   Canuto, V. M., Narlikar, J. V.: Astrophys. J. **236**, 6 (1980)

[23]   Lau, Y. K.: Australian J. of Phys. **38**, 547 (1985)

[24]   Beesham, A.: Int. J. Theor. Phys. **25**, 1295 (1986)

[25]   Abdel-Rahman, A.-M. M.: Gen. Relativ. Gravit. **22**, 665 (1990)

[26]   Kalligas, D., Wesson, P. S., Everitt, C. W. F.: Gen. Relativ. Gravit. **24**, 351 (1992)

[27]   Abdussattar, Vishwarkarma, R. G.: Class. Quant. Grav. **14**, 945 (1997)

[28]   Vishwakarma, R. G.: Gen. Relativ. Gravit. **37**, 1305 (2005)

[29]   Saha, B.: Mod. Phys. Lett. A **20**, 2127 (2005)

[30]   Bali, R., Tinker, S.: Chin. Phys. Lett. **26**, 029802 (2009)

[31]   Klimek, Z.: Nouvo Cimento B 35 249 (1976)

[32]   Murphy, G. L,: Phys. Rev. D **8**, 4231 (1973)

[33]   Fabris, J. C., Goncalves, J. V. B., Ribeiro, R. S.: Gen. Relativ. Grav. **38**, 495 (2006)

[34]   Misner, C. W.: Nature **214**, 40 (1967)

[35]   Padmanabhan, T., Chitre, S. M.: Phys. Lett. A **120**, 433 (1987)

[36]   Singh, C. P.: Gravitation and Cosmol **15**, 381-390 (2009)

[37]   Meng, X., Ma, Z.: Eur. Phys. J. C **72**:2053 (2012)

[38]   Baghel, P. S., Singh, J. P.: Research in Astron Astrophys. **11,** 1457 (2012)

[39]   Das, K., Ali, N.: Natl. Acad. Sci. Lett. **37**, 173 (2014)

[40]    Santos, N. O., Dias, R. S., Banerjee, A.: J. Math. Phys. **26**, 878 (1985)

[41]    Arbab, A. I.: Gen. Relativ. Gravit. **30**, 1401 (1998)

[42]    Beesham, A., Ghosh, S. G., Lombard, R. G.: Gen. Relativ. Gravit. 32, 471 (2000)

[43]    Misner, C. W.; Astrophys. J. **151**, 431 (1868)

[44]    Chimento, L. P.: Phys. Rev. D **69**, 123517 (2004)

[45]    Land, K., Magueijo, J. 2005, Phys. Rev. Lett. **95**, 071301.

[46]    Colay, A. A.: Gen. Relativ. Gravit. **22**, 241 (1990)

[47]    Singh, T., Chaubey, R.: Pramana J. Phys. **67**, 415 (2006)

[48]    Singh, T., Chaubey, R.: Pramana J. Phys. **68**, 721 (2007)

[49]    Singh, J. P., Baghel, P. S.:  EJTP **6**, 85 (2009)

[50]    Singh, J. P., Baghel, P. S.:  Int. J. Theor. Phys. **49**, 273 (2010)

[51]    Yadav, A. K.: Chin. Phys. Lett. **29**, 079801 (2012)

[52]    Pradhan, A., Yadav, L., Yadav, A. K.: Czch. J. Phys. **54**, 487 (2004)

[53]    Singh, C. P., Ram, S., Zeyauddin, M.: Astrophys. Space Sci. **315**, 181 (2008)

[54]    Ellis, G. F. R., Madsen, M.: Class. Quantum Grav. **8**, 667 (1991)

[55]    Hawking, S. and Ellis, G. F. R.: The Large Scale Structure of Space time, p. 88, Cambridge University Press (1975)

[56]    Banerjee, N., Das, S.: Gen. Relativ. Gravit. **37**, 1695 (2005)

[57]    Singha, A. K., Debnath, U.: Int. J. Theor. Phys. **48**, 351 (2009)

[58]    Singh, J. P.: Astrophys. Space Sci. **318**, 103 (2008)

[59]    Singh, J. P.: Int. J. Theor. Phys. **48**, 2041 (2009)

[60]    Bali, R., Singh, P.: Int. J. Theor. Phys. **51**, 772 (2012)

[61]    Starobinskii, A. A.: Sov. Astron. Lett. **7**, 36 (1981)

[62]    Weinberg, S.: Gravitation and Cosmology, Wiley, New York (1972)

[63]   Sahni, V., Saini, T. D., Starobinsky, A. A. Alam, U.: JETP **77**, 201 (2003)

[64]   Alam, U., Sahni, V., Saini, T. D., Starobinsky:Mon. Not. R. Astron. Soc. **344**, 1057 (2003)

[65]   Knop, R. A., et al.: Astrophys. J. **598**, 102 (2003)

[66]   Tegmark, M., et al.: Phys. Rev. D **69**, 103501 (2004)

[67]   Spergel, D. N., et al.: WAMP three years results: Implications for cosmology, preprint, astro-ph/0603449 (2006)

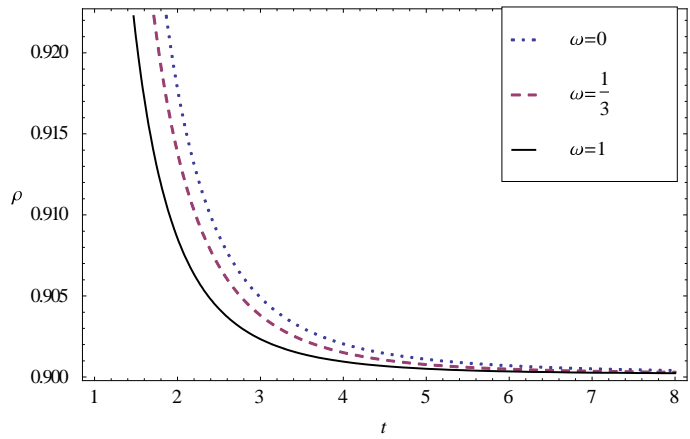

Fig.1 Variation of matter density $\rho$ with cosmic time $t$ by taking $m=0.3$, $n=1.5$, $\zeta_0=0.3$, $\zeta_1=0.4$, $\zeta_2=0.5$ and $k_2=0.5$.

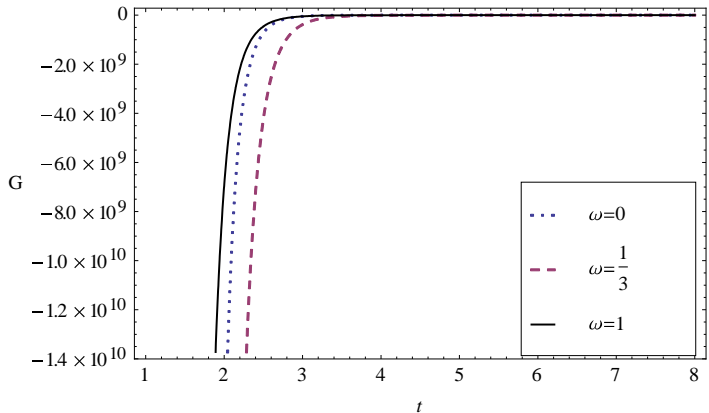

Fig.2 Variation of gravitational constant $G$ with cosmic time $t$ by taking $m=0.3$, $n=1.5$, $\zeta_0=0.3$, $\zeta_1=0.4$, $\zeta_2=0.5$, $k_1=0.2$ and $k_2=0.5$.

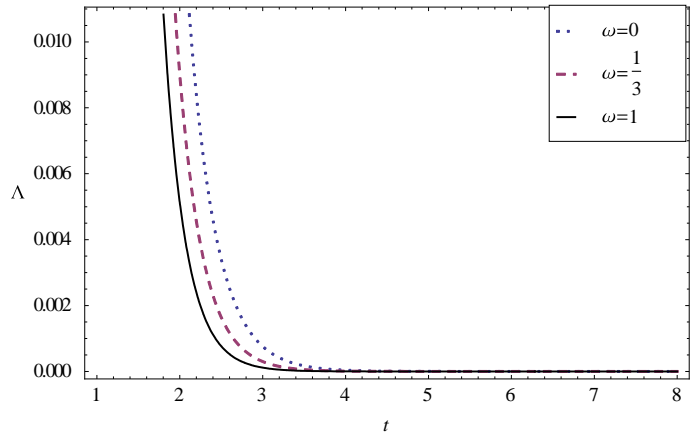

Fig.3 Variation of Cosmological term-$\Lambda$ with cosmic time $t$ by taking $m=0.3$, $n=1.5$, $\zeta_0=0.3$, $\zeta_1=0.4$, $\zeta_2=0.5$, $k_1=0.2$ and $k_2=0.5$.

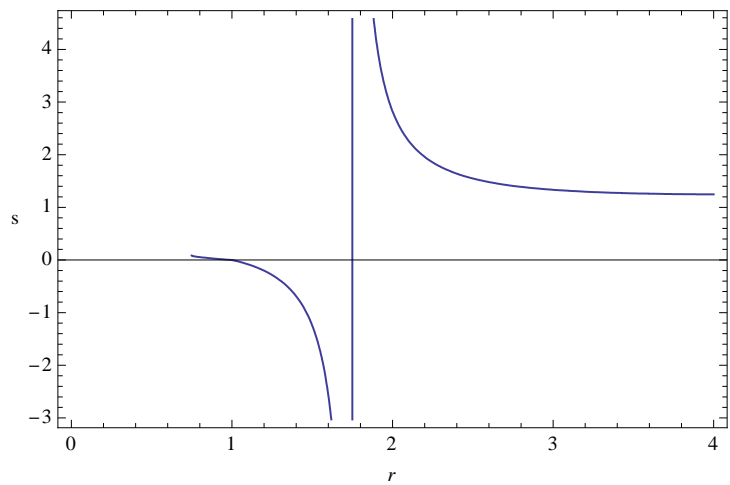

Fig.4 Variation of $s$ against $r$ by taking $n=2$.